\documentclass[twocolumn,showpacs,preprintnumbers,amsmath,eqsecnum, amssymb,prd,superscriptaddress]{revtex4-1}

\usepackage{amsmath,amssymb,amsthm,color}
\usepackage{graphicx}
\usepackage{dcolumn}
\usepackage{subcaption}
\usepackage{commath}

\newcommand {\beq}{\begin{eqnarray}}
\newcommand {\eeq}{\end{eqnarray}}

\newcommand{\dspace}{\ensuremath{d}}
\newcommand{\dez}{\ensuremath{\Delta_0}}
\newcommand{\fext}{\phi}
\newcommand{\NN}{\mathcal{N}}
\newcommand{\hypgeom}{{}_2F_1}
\newcommand{\sfrac}[2]{{\textstyle\frac{#1}{#2}}}
\newcommand{\half}{\sfrac{1}{2}}

\begin{document}

\preprint{CALT-TH 2015-053, IPMU 15-0184}

\title{Reflections on Conformal Spectra}

\author{Hyungrok Kim}
\affiliation{Walter Burke Institute for Theoretical Physics, Caltech, Pasadena, California 91125, USA}
\affiliation{School of Natural Sciences, Institute for Advanced Study, Princeton, New Jersey 08540, USA}

\author{Petr Kravchuk}

\affiliation{Walter Burke Institute for Theoretical Physics, Caltech, Pasadena, California 91125, USA}
\affiliation{School of Natural Sciences, Institute for Advanced Study, Princeton, New Jersey 08540, USA}

\author{Hirosi Ooguri}

\affiliation{Walter Burke Institute for Theoretical Physics, Caltech, Pasadena, California 91125, USA}
\affiliation{School of Natural Sciences, Institute for Advanced Study, Princeton, New Jersey 08540, USA}
\affiliation{Kavli Institute for the Physics and Mathematics of the Universe (WPI), University of Tokyo, Kashiwa 277-8583, Japan}


\begin{abstract}
We use modular invariance and crossing symmetry of conformal field theory
to reveal approximate reflection symmetries in
the spectral decompositions of the partition function in two dimensions in the limit of large central charge
and of the four-point function
in any dimension in the limit of large scaling
dimensions $\Delta_0$ of external operators. We use these symmetries to motivate universal upper bounds on the spectrum and the operator product expansion coefficients, which we then derive by independent techniques.
Some of the bounds for four-point functions are valid for finite $\Delta_0$ as well as for large $\Delta_0$. We discuss a similar symmetry in a large spacetime dimension limit. Finally, we comment on the analogue of the Cardy formula and sparse light spectrum condition for the four-point function.

\end{abstract}

\maketitle

\section{Introduction}

Modular invariance and crossing symmetry  relate
ultraviolet and infrared properties of conformal field theory and impose strong constraints on
its energy spectrum and operator product expansion (OPE).
In two dimensions, the partition function,
\begin{equation}
 Z(\tau) = {\rm tr} \ q^{L_0-\frac{c}{24}}\ \bar{q}^{\bar{L}_0-\frac{c}{24}} ,
\label{partition}
\end{equation}
is invariant under the modular transformation, $\tau \rightarrow -1/\tau$, where $q = e^{2\pi i \tau}$ and $\tau$ is the torus modulus.
In any number of dimensions, a four-point function on the sphere,
\begin{equation}
 G(x) =\langle 0 | \phi(\infty) \phi(1) \phi(x) \phi(0) | 0 \rangle  ,
\label{fourpoint}
\end{equation}
is invariant under the crossing transformation, $x \rightarrow 1 - x$, where $x$ is the Dolan-Osborn coordinate \cite{Dolan:2000ut}.
The use of modular invariance was initiated in  \cite{Cardy:1986ie}.
The conformal bootstrap program to exploit crossing symmetry was pioneered in \cite{Ferrara:1973yt, Polyakov:1974gs},
was developed further in two dimensions starting with \cite{Belavin:1984vu}, and is currently undergoing a renaissance in higher
dimensions starting with \cite{Rattazzi:2008pe}.

The quintessential application
of modular invariance is the Cardy formula \cite{Cardy:1986ie},
which describes the spectral density for a large scaling dimension $\Delta$ with a fixed value
of the central charge $c$.  In \cite{Pappadopulo:2012jk}, crossing symmetry was used  to estimate the spectral density
weighted by the OPE coefficients, for large $\Delta$ with a fixed value of the scaling
dimension  $\Delta_0$ of the external operator $\phi$ in
(\ref{fourpoint}).

In this paper, we will study the different limits:
\begin{equation}
\Delta, c \rightarrow \infty , ~~ {\rm with} ~ \Delta/c:\ {\rm fixed} ,
\label{climit}
\end{equation}
for the partition function,
\begin{equation}
 \Delta, \Delta_0 \rightarrow \infty , ~~ {\rm with} ~ \Delta/\Delta_0 :\ {\rm fixed} ,
\label{deltalimit}
\end{equation}
and
\begin{equation}
	\Delta,\Delta_0,\dspace\rightarrow\infty, ~~ {\rm with} ~ \Delta/\dspace,\,\Delta_0/\dspace :\ {\rm fixed},
\end{equation}
for the four-point function. Here $\dspace$ is the spacetime dimension.

The limit (\ref{climit}) for the partition function was considered in \cite{Hartman:2014oaa},
where it was shown that the Cardy formula holds
for $\Delta > c/6$  under a certain condition on light spectrum,
strengthening the result of \cite{Cardy:1986ie}, which held only in the limit $\Delta \gg c$. In this paper, we will describe an approximate symmetry of spectral decomposition of the partition function, which can be used to motivate this result. Moreover, this symmetry  suggests some bounds for the spectral density, which we derive by independent techniques.
We employ a similar approach to study the limit (\ref{deltalimit}) of the four-point function
to derive properties of the spectral density weighted by the OPE coefficients as a function of
$\Delta$. This approach proves to be universal and we apply it also to the case of large spacetime dimension.

\section{Results}
\subsection{Partition function}
\label{sec:partresults}
To study the partition function in two dimensions, we will use the following simplified expression:
\begin{equation}
Z(\tau)  = \int_0^\infty q^{\Delta - \frac{c}{12}} n(\Delta) d \Delta ,
\label{partitionlimit}
\end{equation}
where $n(\Delta)$ is the density of conformal primary states with scaling dimension $\Delta$.
This formula ignores contributions from Virasoro descendants, which will turn out to be subleading in $1/c$ in what follows. Another interpretation is that $n(\Delta)$ is the density of all states, not just the primaries, in which case the above formula is valid literally. The spins of primary states are not visible when $q$ is real
and $\tau$ is pure imaginary, which we will assume throughout the paper.

Our basic observation is that modular invariance $Z(\tau) = Z(-1/\tau)$ implies the
following approximate reflection symmetry in the space of scaling dimension $\Delta$:
\begin{equation}
  \bar\omega_\tau (\Delta) \simeq  \bar\omega_{-1/\tau}\left( (1+|\tau|^2) \frac{c}{12} - |\tau|^2 \Delta \right)\times |\tau|^2,
\label{reflectiontau}
\end{equation}
where $\bar\omega_\tau (\Delta)$ is defined by,
\begin{align}
  \omega_\tau (\Delta) &= \frac{1}{Z(\tau)}q^{\Delta - \frac{c}{12}} n(\Delta),\\
  \bar\omega_\tau (\Delta) &= K_c(\Delta)*\omega_\tau(\Delta),
\label{significancetau}
\end{align}
and $*$ denotes convolution, $(f*g)(x)=\int f(x-y)g(y)dy$. Here the kernel $K_c$ smears the integrand of \eqref{partitionlimit} over the interval of size $\varepsilon$, $\sqrt c\ll \varepsilon\ll c$. Note however that $K_c$ decays rather slowly outside of this interval -- see section III.A.1.
With this definition, $\bar\omega_\tau$ measures the significance of $\Delta$ in the partition function
averaged over the small interval of the size $\varepsilon$ to smooth out the sum of delta-functions in $n(\Delta)$.
Since $\Delta$ is bounded below by $0$ in any unitary theory,
$\bar\omega_\tau(\Delta)$ approximately vanishes for $\Delta \lesssim -\varepsilon$. The reflection symmetry (\ref{reflectiontau}) maps this to,
\begin{equation}
 \bar\omega_\tau(\Delta \gtrsim \Delta_\tau) \simeq 0,
\end{equation}
where the edge $\Delta_\tau$ is given by,
\begin{equation}
 \Delta_\tau = \left(1 + \frac{1}{|\tau|^2}\right) \frac{c}{12}.
\label{saddletau}
\end{equation}
We can estimate how fast the integrand of \eqref{partitionlimit} decays above this threshold $\Delta > \Delta_\tau$, $|\tau|<1$, as,
\begin{equation}
\int_{\Delta}^\infty  \omega_\tau (\Delta')d\Delta' \leq \frac{2}{1 + T_{2k_0+1}\left(\frac{\Delta-\Delta_\tau/2}{\Delta_\tau/2}\right)},
\label{abovethreshold}
\end{equation}
where $T_{2k_0+1}(x)$ is the degree $(2k_0+1)$ Chebyshev polynomial of the first kind
and $k_0$ is chosen so that $k_0 \ll \sqrt{c}$. In the limit of $c\rightarrow \infty$, the half decay width of the right hand side is $\sim c/k_0\gg \sqrt c$.

Of course, from Cardy formula one expects exponential rather than polynomial decay, but this formula shows the specific threshold value $\Delta_\tau$, beyond which there can be no dominant contribution to $Z(\tau)$. From the discussion in \cite{Hartman:2014oaa} it follows that there exist theories which essentially saturate this bound, i.e. for which the integral \eqref{partitionlimit} is dominated by states at $\Delta_\tau$.

This happens in theories satisfying the sparse light spectrum condition, defined in \cite{Hartman:2014oaa} as,
\begin{align}
n(\Delta')&  \lesssim e^{2\pi\Delta}, \nonumber\\
&{\rm for} ~ 0 \leq \Delta < c/12,
\label{sparselightspectrumcondition}
\end{align}
where the inequality should be understood in an averaged sense.
The essence of this condition is that the partition function for the low temperature
phase $|\tau| > 1$ is dominated by the vacuum state
(in particular, the maximum of $\omega_\tau(\Delta)$
is at $\Delta=0$). In this case, the reflection symmetry shows that
the maximum of $\omega_\tau(\Delta)$
jumps to the edge $\Delta_\tau$ in the high temperature
phase  $|\tau| < 1$, and gives a prediction on the value of this maximum. With $\tau$ changing in the high-temperature phase the maximum at $\Delta_\tau$ scans through the region $\Delta>c/6$, allowing one to obtain information on $n(\Delta)$ in this region.
Rigorous microscopic estimates were made in \cite{Hartman:2014oaa}, and the resulting Cardy-like formula is
\begin{equation}
 \bar{n}(\Delta) \simeq \exp\left( 2\pi \sqrt{\frac{c}{3}\left(\Delta - \frac{c}{12}\right)}+O(c^\alpha)\right) ,
\label{Cardy}
\end{equation}
for $\Delta>c/6$ and the average density of states,
\begin{equation}
\bar{n}(\Delta)
=  \frac{1}{\varepsilon'} \int_\Delta^{\Delta + \varepsilon'} n(\Delta')d\Delta',
\end{equation}
with $\varepsilon'\sim c^\alpha$, $1/2<\alpha<1$.

\subsection{Four-point function}
In this section we consider the four-point function of identical scalar operators of scaling dimension $\Delta_0$. We insert the four operators on one two-dimensional plane, which we identify with the complex plane of variable $x$. We insert three scalars at $0,1,\infty$ and the fourth scalar at the Dolan-Osborn coordinate $x$.
This four-point function (\ref{fourpoint}) can be expressed as a sum of the spectral density weighted by
the OPE coefficients and the conformal block ${\mathcal F}_{\Delta, \ell}(x)$ for the scaling dimension $\Delta$ and the spin $\ell$, see e.g. \cite{Pappadopulo:2012jk}.
Here and throughout the paper, we assume that the coordinate $x$ is real and $0 < x < 1$.

As a by-product of our work, we find an expression for ${\mathcal F}_{\Delta, \ell}(x)$ for general $\ell$ in the scaling limit (\ref{deltalimit}), when
external operators are identical scalars.
In Appendix A, we will solve the fourth order differential equation derived in \cite{Hogervorst:2013kva} for the conformal block
to show, for $x<1$,
\begin{align}
{\mathcal F}_{\Delta, \ell}(x)  \simeq \ & \rho^\Delta \left(1 - \frac{\rho^2}{16}\right)^{-\frac{d}{2} + \kappa(\Delta, \ell,\rho)} \nonumber \\
         & ~~~ \times \left( 1 + O(1/\Delta)\right) ,
\label{approximateblock}
\end{align}
where $\rho$ is the radial coordinate,
\begin{equation}
 \rho = \frac{4x}{(1 + \sqrt{1-x})^2} .
\label{whatisrho}
\end{equation}
 introduced in \cite{Pappadopulo:2012jk} and discussed further in
 \cite{Hogervorst:2013sma}. Note that this approximation breaks down when $x\to 1$. This should be kept in mind when interpreting the formulas below. In general the results of this section apply to the limit $\Delta_0\to\infty$ with $x$ kept fixed.
In this limit, spin dependence of the conformal block is only through the exponent $\kappa(\Delta, \ell,\rho)$, which behaves as,
\begin{align}
\kappa(\Delta, \ell,\rho) & \rightarrow 0, ~~ (\Delta-\ell\sim \Delta ) , \\
    &\rightarrow \frac{1}{2}, ~~ (\Delta = \ell + d-2:  {\rm unitarity~bound}) .\nonumber
\end{align}
Here in the first case $\ell$ can be on the order of $\Delta$, but has to stay away from the unitarity bound. Between the two cases $\kappa$ can acquire $\rho$ dependence. However, the results in the two regimes suggest that the factor $(1 - \rho^2/16)^{-d/2 + \kappa(\Delta, \ell, \rho)}$
in the conformal block (\ref{approximateblock}) is
altogether negligible in the large $\Delta$ analysis in this paper, just as Virasoro descendants are negligible in the
partition function as in (\ref{partitionlimit}).
Thus, we can express the four-point function in the scaling limit (\ref{deltalimit}) as,
\begin{equation}
G(x)  = \int_0^\infty \rho^{\Delta} x^{-2\Delta_0} g(\Delta) d\Delta ,
\label{fourpointlimit}
\end{equation}
where  $g(\Delta)$ is the spectral density weighted by the square of
the OPE coefficients, which is non-negative when $\phi$'s are identical. One can of course keep this subleading factor in what follows without affecting the conclusions.
Note that though we made no assumptions on the spins of the intermediate states,
the spectral decomposition of $G(x)$ is blind to them for real $x$ and large scaling dimensions.

One can also view \eqref{fourpointlimit} as an exact expansion, in which we have discarded the structure of conformal multiplets and treat primary and descendant operators on equal footing. This is the radial coordinates expansion of \cite{Pappadopulo:2012jk,Hogervorst:2013sma}. Below we also consider another kind of ``descendant'' expansion, which corresponds to a different choice of coordinates.

Since the spectral decomposition of the four-point function (\ref{fourpointlimit}) is similar to that of
the partition function (\ref{partitionlimit}) in these limits, crossing symmetry $G(x) = G(1-x)$ implies
a similar reflection symmetry in $\Delta$. Let us introduce the ``branching ratio'' of $\phi(x) \times \phi(0)$ turning into operators of dimension $\Delta$,
\begin{align}
\gamma_x(\Delta) &= \frac{1}{G(x)} \rho^{\Delta} x^{-2\Delta_0} g(\Delta),\\
\bar\gamma_x(\Delta) &= K_{\Delta_0}(\Delta)*\gamma_x(\Delta),
\label{branchingx}
\end{align}
with $K_{\Delta_0}$ averaging over intervals of the size $ \sqrt{\Delta_0}\ll \varepsilon \ll \Delta_0$.
In terms of this quantity, the approximate reflection symmetry is expressed as,
\begin{equation}
   \bar\gamma_x(\Delta) \simeq \bar\gamma_{1-x}\left( \frac{1}{\sqrt{x}}\left( 2\Delta_0 - \sqrt{1-x}\Delta\right) \right)\sqrt{\frac{1-x}{x}}.
\label{reflectionx}
\end{equation}
The reflection of $\bar\gamma_x(\Delta \lesssim -\varepsilon ) = 0$ is then,
\begin{equation}
 \bar\gamma_x(\Delta \gtrsim \Delta_x) \simeq 0,
\end{equation}
where the edge $\Delta_x$ is given by,
\begin{equation}
 \Delta_x = \frac{2}{\sqrt{1-x}} \Delta_0.
\label{saddlex}
\end{equation}
As in the case of the partition function (\ref{abovethreshold}), we can estimate how fast $\gamma_x(\Delta)$ decays above the threshold
$\Delta > \Delta_x$, $x>1/2$ as,
\begin{equation}
 \int_\Delta^\infty
\gamma_x(\Delta') d\Delta' \leq \frac{2}{1 + T_{2k_0+1}\left(\frac{\Delta-\Delta_x/2}{\Delta_x/2}\right)},
\label{chebyshev4pt}
\end{equation}
with $k_0 \ll \sqrt{\Delta_0}$. Note that the half-decay width is $\sim \Delta_0/k_0\gg\sqrt{\Delta_0}$. This can be compared to the conformal block expansion of the correlation function of the generalized free field,
\begin{equation}
	G(x)=\frac{1}{x^{2\Delta_0}}+\frac{1}{(1-x)^{2\Delta_0}}+1,
\end{equation}
which can be shown, as long as $x$ is away from $1$, to have a saddle point at $\Delta=\Delta_x$ of width $\sim \sqrt{\Delta_0}$. In this order-of-magnitude sense the bound \eqref{chebyshev4pt} is almost saturated.

We can also perform the ``descendant'' expansion in the standard coordinates described in the beginning of this section (see e.g. \cite{Pappadopulo:2012jk}), again treating primary and descendant operators on equal footing,
\begin{equation}
	G(x)=\int_0^\infty x^{\Delta-2\Delta_0} g^{(s)}(\Delta)d\Delta,
\end{equation}
where we added the superscript ${}^{(s)}$ to $g(\Delta)$ to
note the fact that we are expanding $G(x)$ in what we will henceforth call ``scaling blocks''. We use a similar notation for branching ratios $\gamma^{(s)}_x,\,\bar\gamma^{(s)}_x$.
All of the above results also hold in this case, with the modification that now
\begin{equation}
	\Delta_x = \frac{2}{1-x},
\end{equation}
the reflection relation is
\begin{equation}
\bar\gamma_{x}^{(s)}(\Delta) \simeq \bar\gamma_{1-x}^{(s)}\left( \frac{2}{x} \Delta_0 - \frac{1-x}{x} \Delta\right)\frac{1-x}{x}.
\end{equation}

\subsubsection{Finite-$\Delta_0$ bounds}

So far, our statements have been in the limit (\ref{deltalimit}) of large  $\Delta$ and $\Delta_0$. In the case of the scaling block decomposition of four-point function,
we can derive inequalities which are valid at finite $\Delta$ and $\Delta_0$.
For example,  for $2<4\Delta_0 < \Delta$,
\begin{equation}
 \int_\Delta^\infty
\gamma_{1/2}^{(s)}(\Delta') d\Delta' \leq
\frac{1}{1
+\frac{\Gamma(\Delta - 2\Delta_0+1)\Gamma(2\Delta_0)}
{\Gamma\left(\frac{\Delta+3}{2}\right)\Gamma\left(\frac{\Delta-1}{2}\right)}},
\label{finitebound}
\end{equation}
where
\begin{equation}
	\gamma_{x}^{(s)}(\Delta)=\frac{1}{G(x)}x^{\Delta-2\Delta_0}g^{(s)}(\Delta).
\end{equation}
Note that this bound also implies a bound on individual delta-function contributions to $g^{(s)}$, since they are all positive. If we keep $\Delta_0$ finite and take $\Delta \rightarrow \infty$, this inequality becomes,
\begin{equation}
 \int_\Delta^\infty
\gamma_{1/2}^{(s)}(\Delta') d\Delta' \leq \sqrt{2\pi} \frac{\Delta^{2\Delta_0 - \frac{1}{2}}}{2^\Delta \Gamma(2\Delta_0)}.
\end{equation}

In this limit, this inequality is stronger than the asymptotic bound of \cite{Pappadopulo:2012jk},
\begin{equation}
\int_\Delta^\infty
\gamma_{1/2}^{(s)}(\Delta') d\Delta'\lesssim \frac{2^{-2\Delta_0}}{G(1/2)}\frac{\Delta^{2\Delta_0}}{2^\Delta \Gamma(2\Delta_0+1)}.\label{opeconvbound}
\end{equation}
However, the Cardy-like asymptotic of \cite{Pappadopulo:2012jk},
\begin{equation}
\int_0^\Delta
 g^{(s)}(\Delta') d\Delta' \sim \frac{\Delta^{2\Delta_0 }}{\Gamma(2\Delta_0+1)},
\end{equation}
suggests by differentiation that one can expect the stronger convergence rate of
\begin{equation}
\int_\Delta^\infty
\gamma_x^{(s)}(\Delta') d\Delta' \propto \Delta^{2\Delta_0-1}2^{-\Delta}.
\end{equation}
While \eqref{finitebound} is weaker than this expectation, it has the advantage that it is rigorous and holds for finite $\Delta$ and $\Delta_0$.

In fact, the method we use for proving this bound is quite general and can be used for construction of finite $\Delta$ and $\Delta_0$ analytic bounds for \eqref{fourpointlimit} as well. We have checked that these bounds are asymptotically at least as strong as those of \cite{Pappadopulo:2012jk}, still having the advantage of being valid for finite values of $\Delta$. Given the improvement of \eqref{finitebound} over \eqref{opeconvbound}, one might expect that an improvement is possible for \eqref{fourpointlimit} as well. We hope to return to this question in future.

So far, we did not assume that the four-point function is dominated by a saddle point.  If we make this assumption,
our results have simple explanation. Let the location of the saddle point in the expansion of $G(x)$ be $\Delta(x)$, which has to obey the reflection relation imposed by crossing symmetry,
\begin{equation}
	\frac{\Delta(x)-2\Delta_0}{x}=-\frac{\Delta(1-x)-2\Delta_0}{1-x}.
	\label{saddlereflection}
\end{equation}
This is most easy to see if we note that $\Delta(x)-2\Delta_0=\frac{\partial \log G(x)}{\partial \log x}$ and apply the crossing relation $G(x)=G(1-x)$.
In unitary theory $\Delta(x)\geq 0$, which implies, by the above relation,
\begin{equation}
	\Delta(x)\leq \Delta_x=\frac{2\Delta_0}{1-x}.
	\label{saddlebound}
\end{equation}

\subsubsection{Cardy formula}

An analogue of the sparse light spectrum condition (\ref{sparselightspectrumcondition}) for the four point
function can be introduced, namely,
\begin{align}
g^{(s)}	(\Delta')\lesssim  2^\Delta , \nonumber\\
&{\rm for} ~ 0 <\Delta < 2\Delta .
\end{align}
Again, this should be understood in some averaged sense, such that this condition would imply that
the four-point function for $|x| < 1/2$ is dominated by the vacuum state.
Then, by the reflection symmetry,
the maximum of $\gamma_x(\Delta)$ jumps to the edge $\Delta_\tau$ for $|x| > 1/2$.
This, exactly as in the case of the partition function, can be translated into a statement on $g^{(s)}(\Delta)$, which
reads, for $\Delta>\Delta_{1/2}=4\Delta_0$,
\begin{multline}
	\bar g_s(\Delta)=\exp\left[-\Delta \log\left(1-\frac{2\Delta_0}{\Delta}\right)+\right.
	\\\left.+2\Delta_0\log \left(\frac{\Delta}{2\Delta_0}-1\right)+O(\Delta_0^\alpha)\right],
	\label{4ptcardy}
\end{multline}
where $1/2<\alpha<1$, and $\bar g^{(s)}$ is $g^{(s)}$ integrated on the scale $\delta\sim \Delta_0^\alpha$.

\subsection{Four-point function in large spacetime dimension}
\label{sec:doublescaleresults}

So far we have only discussed the limits where the operators considered were heavy compared to any other scales we had, and in particular far away from the unitarity bounds. Some interesting phenomena happen near unitarity bounds, such as that a scalar field has to become free as its scaling dimension is pushed toward the bound. In this section we consider a limit in which we take not only the scaling dimension of the external scalars, but also the number of spacetime dimensions $\dspace$ to be large. In fact, when the number of spacetime dimensions is taken to be large, the unitarity bounds force all the operators to become heavy. We are then able to apply the same methods as before, but now to all operators in the theory.

Recall that the unitarity bounds are
\begin{align}
\Delta&\geq \frac{\dspace-2}{2}\sim\frac{\dspace}{2}&\text{for non-identity scalars},\\
\Delta&\geq \ell+\dspace-2\sim\ell+\dspace&\text{for operators with spin},
\end{align}
and thus the natural limit is the double scaling $\dez\sim\dspace\to\infty$. In this limit we can see the gap between the identity and the lightest allowed scalars and the difference between the lightest scalars and the lightest spin operator. Appearance of these features means that now we have to distinguish several classes of operators.

It turns out that for us there is no difference between spin and scalar operators, since on the real line $x=\bar x$ the conformal blocks in a large number of spacetime dimension do not depend on spin (for details on conformal blocks see Appendix~\ref{app:doublescale}). However, the gap above the identity is important and the identity operator has to be treated separately.

As mentioned before, we can apply almost the same methods as we used in other limits. A new feature is that the duality relation is now non-linear and is not as pleasant to manipulate as in the above discussions. However, it carries more information, since we are now able to take our external scalars close to the unitarity bound.

Let us introduce the duality relation. We state it in the following form,
\begin{equation}
	\Lambda_x(\Delta)= -\Lambda(\Delta')_{1-x}.
\end{equation}
This has to be understood as an implicit relation between the symmetry-related scaling dimensions $\Delta$ in the conformal block expansion at $x$ and $\Delta'$ at $1-x$. Here $\Lambda_x$ is given by
\begin{equation}
	\Lambda_x(\Delta)=\frac{1}{\dez}\frac{\partial \log x^{-2\dez}F_\Delta(x)}{\partial x},
\end{equation}
and $F_\Delta$ is the spin-independent conformal block. The explicit form of $\Lambda_x$ is cumbersome, but is straightforwardly obtained from \eqref{eq:doublescalelog} for $\Delta>0$. For the identity operator $F_0(x)=1$, and so we get $\Lambda_x=-2/x$.
One can easily obtain the range of $\Lambda_x$ corresponding to the unitary range $\Delta\in\{0\}\cup[d/2,+\infty)$. It is given by
\begin{equation}
	\Lambda_x\in\left\{-\frac{2}{x}\right\}\cup\left[\frac{1}{2\delta_0 x(1-x)}-\frac{2}{x},+\infty\right),
\end{equation}
where $\delta_0=\dez/\dspace$. Now, let us apply the duality relation to this range -- in this way we will obtain the allowed range for the saddle point in the conformal block decomposition. The result is, in terms of $\Lambda$,
\begin{multline}
	\Lambda_x\in\left\{-\frac{2}{x}\right\}\cup\left\{\frac{2}{1-x}\right\}\\
	\cup\left[\frac{1}{2\delta_0 x(1-x)}-\frac{2}{x},-\frac{1}{2\delta_0 x(1-x)}+\frac{2}{1-x}\right].
\end{multline}

\begin{figure*}[t]
	\centering
	\begin{subfigure}[b]{0.45\textwidth}
	\includegraphics[width=\textwidth]{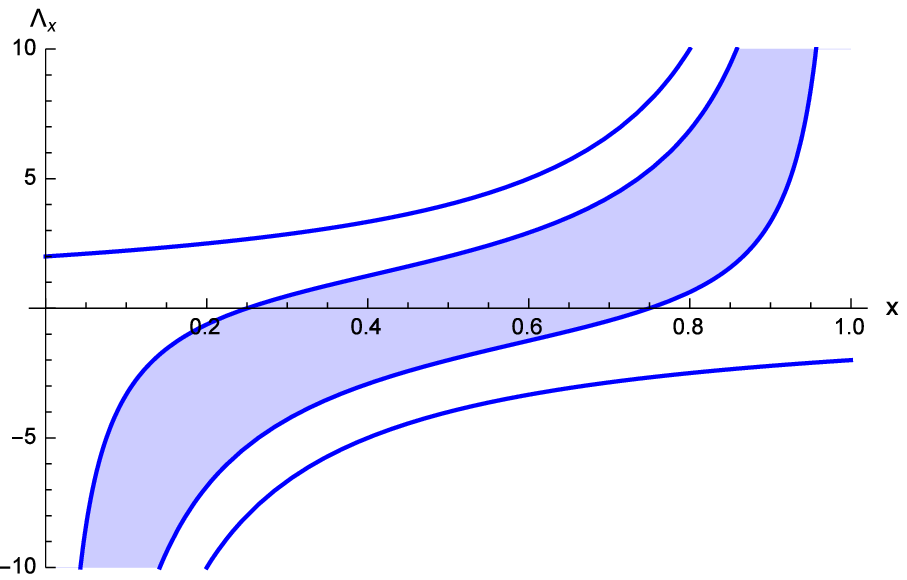}
	\caption{$\delta_0=1$}
	\label{fig:lambda_1}
	\end{subfigure}
	~
	\begin{subfigure}[b]{0.45\textwidth}
	\includegraphics[width=\textwidth]{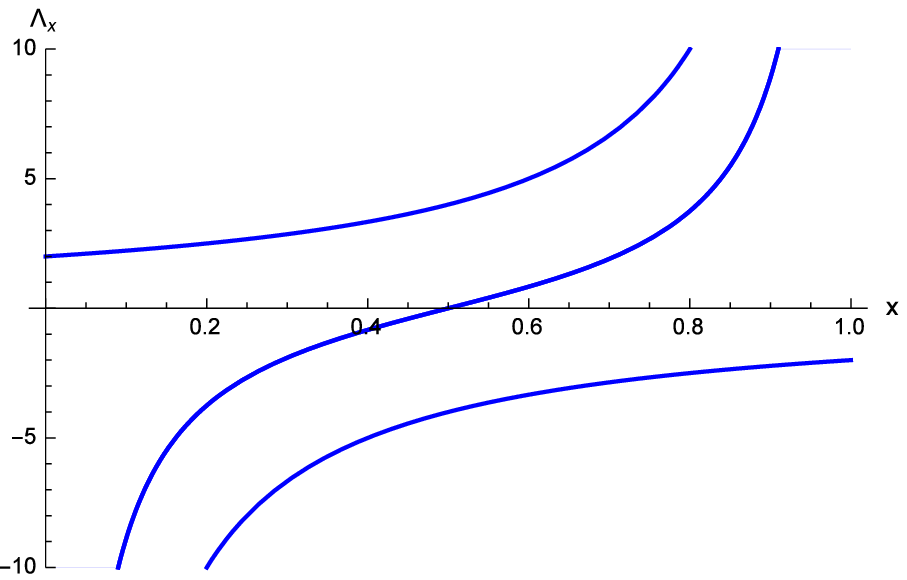}
	\caption{$\delta_0=1/2$}
	\label{fig:lambda_05}
	\end{subfigure}
	\caption{Allowed range for $\Lambda_x$ as a function of $x$ for $\delta_0=1$ and $\delta_0=1/2$.}
	\label{fig:lambda}
\end{figure*}

This range is plotted in Fig.~\ref{fig:lambda}. The case of $\delta_0=1$ is generic and is shown in Fig.~\ref{fig:lambda_1}. As the external scalar gets heavier, $\delta_0$ gets larger and the range fills the region between the curves corresponding to the identity operator and its dual image.

An interesting thing happens as $\delta_0$ approaches the unitarity bound $1/2$, Fig~\ref{fig:lambda_05}. The allowed range for $\Lambda_x$ shrinks into three points. This is the manifestation of the fact that a scalar at the unitarity bound has to be free. Let us remind the reader of the reasoning. The unitarity bound $\Delta\geq (d-2)/2$ expresses non-negativity of the norm of a descendant of $\fext$, which thus becomes null at the unitarity bound. This implies that $\fext$ satisfies the free field equation of motion $\Delta\fext=0$ as an operator equation, and all the correlation functions of $\fext$ are harmonic away from singularities. Then one can take for example the four point function of $\fext$ and subtract the free field four point function. The result $G'_4$ is still harmonic and the OPE limits imply that it has singularities weaker than those of free field, $1/|x|^{d-2}$. But $1/|x|^{d-2}$ is the weakest singularity a harmonic function can have. Thus, $G'_4$ is harmonic everywhere, tends to zero at infinity and is therefore $0$. So the four point function of $\fext$ is that of the free field, which in turn implies that the $\fext\fext$ OPE is also free.

Note that the above argument explicitly imposes the equation of motion of $\fext$ on the four point function. It is not a priori obvious that the crossing equation for this four point function alone should also imply that $\fext$ must be free at the unitarity bound. However, it seems to be the case as the numerical results suggest (e.g. \cite{Rattazzi:2008pe} in four spacetime dimensions). From our perspective, it is true as long as one excludes the middle curve in Fig.~\ref{fig:lambda_05}. If this is done, then duality at $x=1/2$ tells us that there are to equally important saddle points, and for other values of $x$ one of them dominates, just as in the previous discussion. The resulting behavior is characteristic of the free field, to the accuracy of our approximation.

Section III is devoted to derivations.
Some of technical details are discussed in appendices, including the derivation of (\ref{approximateblock}).

\section{Derivations}

\subsection{Modular Invariance}

\subsubsection{Reflection Symmetry}

Here we discuss the derivation of the reflection symmetry \eqref{reflectiontau}. We do not try to make the derivation very detailed or completely rigorous, since we only use \eqref{reflectiontau} as a heuristic device, and our other derivations are independent of it.

Parametrizing $\tau$ in the partition function as,
\begin{equation}
 \tau = i e^{x-\frac{1}{2}},
\end{equation}
the modular transformation $\tau \rightarrow -1/\tau$ becomes the reflection $x \rightarrow 1 -x$. Therefore,
\begin{equation}
\frac{\partial^{2k+1}}{\partial x^{2k+1}} Z(\tau(x))_{\big| x=1/2} = 0, ~~ k=0,1,2,\cdots ,
\end{equation}
and this can be expressed the integral constraints on $\omega_\tau(\Delta)$ as,
\begin{equation}
  \int_0^\infty \left[ \Delta - \frac{c}{12}\right]^{(2k+1)} \omega_{\tau=i}(\Delta) d\Delta = 0,
\end{equation}
where $\omega_\tau(\Delta)$ is defined by (\ref{significancetau}) and
the bracket symbol $[ \Delta - c/12]^{(2k+1)}$  is defined by,
\begin{align}
  [y]^{(N)} &\equiv e^{2\pi y e^x} \left(-\frac{1}{2\pi}\frac{\partial}{\partial x}\right)^N e^{-2\pi y e^x}{}_{\big| x=0}, \nonumber\\
&= y^N \left( 1 + \frac{N(N-1)}{2y} + \cdots \right) .
\end{align}
When $N \ll \sqrt{|y|}$, we can approximate $[y]^{(N)}$ by the monomial $y^N$. Note that if we use the full Virasoro character instead of $q^{\Delta-c/12}$, this approximation is still valid. It is in this sense in which we said previously that Virasoro descendants are subleading.
Therefore,
 \begin{equation}
  \int_0^\infty \left( \Delta - \frac{c}{12}\right)^{2k+1} \omega_{i}(\Delta) d\Delta \simeq 0,
  \label{approxderiv}
\end{equation}
for $k \ll \sqrt{\Delta}, \sqrt{\Delta_0}$, assuming that the region near $\Delta = c/12$ does
not make a major contribution to the integral, which is consistent with results we will find.
This suggests that $\omega_{i}(\Delta)$
is approximately symmetric under reflection at $\Delta = c/12$:
\begin{equation}
\omega_{i}(\Delta ) \simeq \omega_{i}\left(\frac{c}{6} - \Delta \right) .
\end{equation}
If the dominant contribution came from $c/12$, approximate symmetry like this would be self-evident.

Of course, one cannot expect a literal equality like this -- in the end, we only have a finite number of equations \eqref{approxderiv}. To formulate a more precise statement, let us look at the case of general $\tau$.
For $\tau \neq i$, we have for any $k\geq 0$,
\begin{equation}
	\frac{\partial^{k}}{\partial x^{k}} Z(\tau(x)) = (-1)^k\frac{\partial^{k}}{\partial x^{k}} Z(\tau(1-x)),
\end{equation}
which, with similar approximations, translates into
\begin{multline}
	\int_0^\infty\left[2\pi|\tau|\left(\Delta-\frac{c}{12}\right)\right]^k \omega_\tau(\Delta)d\Delta=\\
	=\int_0^\infty\left[\frac{2\pi}{|\tau|}\left(\frac{c}{12}-\Delta\right)\right]^k \omega_{-1/\tau}(\Delta)d\Delta,
	\label{momentseq}
\end{multline}
for $k\ll\sqrt c$. This is now an equality between some polynomial moments of $\omega_\tau$ and $\omega_{-1/\tau}$, which after some linear changes of arguments and densities $\omega$ can be translated into
\begin{equation}
\int_{-\pi|\tau|c/6}^\infty \lambda^k\omega_\tau'(\lambda)d\lambda=\int_{-\infty}^{\pi c/6|\tau|} \lambda^k\omega_{-1/\tau}'(-\lambda)d\lambda,\label{momentlambda}
\end{equation}
where $\lambda$ is a rescaled version of $\Delta$, and $\omega'$ is the rescaled and renormalized version of $\omega$. We will see below that with $k$ bounded above by $\sqrt c$, the integrals can be restricted to finite intervals of size $\sim c$, up to $1/c$ errors. Then one has an equality of polynomial moments of two functions on finite interval. In other words, their convolutions with any polynomial kernel coincide, provided the degree of the polynomial is bounded by $\sqrt c$. One can then try to pick a delta-like kernel $K'_c(\lambda)$, for example,
\begin{equation}
	K'_c(\lambda)=\left(\frac{l^2-\lambda^2}{l^2}\right)^{k/2},\label{examplekernel}
\end{equation}
where $l$ is twice the size of the interval to which we restrict the integrals in \eqref{momentseq}. Then, restoring the original variables, we have the required claim \eqref{reflectiontau}. Note that this particular delta-like kernel would average over regions of size $\gg c^{3/4}$. One can do better, for details see \cite{Saff}.

\subsubsection{Bound on Tail}

As discussed in Section \ref{sec:partresults}, the reflection symmetry \eqref{reflectiontau} suggests
that $\omega_\tau (\Delta)$ approximately
vanishes for $\Delta > \Delta_\tau$. To understand how good the statement is,
we should estimate an upper
bound on
$\omega_\tau(\Delta)$ when $\Delta$ goes above the threshold $\Delta_\tau$.
At $\tau = i$, the conditions on $\omega_i(\Delta)$ are,
\begin{align}
& \int_0^\infty \omega_{i} (\Delta) d\Delta = 1, \nonumber \\
& \int_0^\infty \left[ \Delta - \frac{c}{12}\right]^{(2k-1)} \omega_{i}(\Delta) d\Delta = 0,
\label{linearconditions}
\end{align}
and,
\begin{equation}
 \omega_i(\Delta) \geq 0.
\end{equation}
What we want to do is to estimate an upper bound on $\omega_i(\Delta)$ at
a particular value $\widehat{\Delta}$ by maximizing the value of
 $\omega_i(\Delta)$ under these conditions. This is a typical linear optimization problem.

Generally speaking, the maximum value (optimal value for the primal problem) of $\vec{c}\cdot \vec{x}$ subject to,
\begin{equation}
A\vec{x} = \vec{b}, ~~{\rm and}~~\vec{x} \geq 0,\label{primalconstr}
\end{equation}
is equal to the minimum value (optimal value for the dual problem) of $\vec{b}\cdot \vec{y}$ subject to
\begin{equation}
A^T \vec{y}\geq \vec{c}.\label{dualconstr}
\end{equation}
This is a statement of the strong duality theorem of linear programming \cite{ConvexOptimization}, which is valid for finite-dimensional vector spaces.
In our case, $\vec{x}$ is an infinite dimensional vector whose entries are values of $\omega_i(\Delta)$ at different values of $\Delta$,
$A$ is a set of integral transforms mapping $\omega_i(\Delta)$ to the left-hand side of (\ref{linearconditions}),
and $\vec{b}=(1,0,0,\cdots)$ as in its right-hand side. Although we still expect the strong duality to hold in our case, we really need only the weak duality, which says that the optimal value for the dual problem (in fact, any feasible value) puts an upper bound on the optimal value of the primal problem. This weaker duality is straightforward to see. Indeed, let $x$ be a solution to \eqref{primalconstr}, then for any $y$ a solution to \eqref{dualconstr} we have
\begin{equation}
	\vec{b}\cdot\vec{y}=\vec{x}\cdot A^{T}\vec{y}\geq \vec{x}\cdot\vec{c}.
\end{equation}

Before discussing what the dual problem is in our case, we first note that maximizing $\omega_i(\widehat\Delta)$ does not make much sense, since $\omega_i$ appears only inside the integrals in the constraint equations, and thus its value at a point is irrelevant unless $\omega_i$ has a delta-function singularity at $\widehat\Delta$. Therefore, it only makes sense to maximize the coefficient of delta-singularity in $\omega_i$ at $\widehat\Delta$.

It is an easy exercise to check that in our case the dual minimization problem then is to minimize $y_0$, subject to
\begin{align}
	P_0(\Delta)&\geq 0,\quad \forall\Delta\geq 0,\\
	P_0(\widehat\Delta)&\geq 1,\label{deltacondition}
\end{align}
where
\begin{equation}
	P_0(\Delta)=y_0+\sum_{k=1}^{\infty}\left[ \Delta - \frac{c}{12}\right]^{(2k-1)} y_k.
\end{equation}
Setting $\Delta=c/12$ we get $y_0\geq 0$, and thus if $P_0(\widehat\Delta)>1$, we can always decrease $y_0$ by dividing $\vec{y}$ by $P_0(\widehat\Delta)$. Thus we may assume $P_0(\widehat\Delta)=1$.

For convenience, we consider $\vec\lambda=\vec y/y_0$, and then the minimum value of $y_0$ is equal to the minimal value of $1/P(\widehat{\Delta})$, where
\begin{equation}
 P(\Delta) = 1 + \sum_{k=1}^\infty  \left[ \Delta - \frac{c}{12}\right]^{(2k-1)} \lambda_k,
\end{equation}
with $\lambda_k$'s being variables,
subject to $P(\Delta) \geq 0$ for all $\Delta$. This is the form of the dual problem most suitable for our purposes. For a different perspective on this problem see \cite{Krein}.

We can find a weaker bound on $\omega_\tau(\Delta)$ by utilizing the conditions (\ref{linearconditions}) for a restricted set of $k$'s,
such as $k=0,1,2,\ldots,k_0$ for some $k_0 \ll c$. Let us first consider the case of $\tau=i$ again.
For $k \ll c$, we can approximate $[ \Delta - c/12]^{(k)}$ by
the monomial $( \Delta - c/12)^{k}$. Our task is then to minimize $1/P_{k_0}(\widehat{\Delta})$, where
\begin{equation}
  P_{k_0} (\Delta) = 1 + \sum_{k=1}^{k_0}  \left( \Delta - \frac{c}{12}\right)^{2k-1} \lambda_k ,
\end{equation}
under the condition $P_{k_0}(\Delta ) \geq 0$ for $\Delta \geq 0$. This is the same problem as maximizing the degree $(2k_0-1)$ odd polynomial,
\begin{equation}
   Q_{k_0}(\Delta)=  \sum_{k=1}^{k_0}  \left( \Delta - \frac{c}{12}\right)^{2k-1} \lambda_k,
\end{equation}
under the condition, $ Q_{k_0}(\Delta) \geq -1$ for $\Delta \geq 0$. Since $Q_{k_0}(\Delta)$ is odd under the reflection
$\Delta \rightarrow c/6 - \Delta$,  within the reflection symmetric interval  $0\leq \Delta \leq c/6$,
$ Q_{k_0}(\Delta) \geq -1$ also implies  $ Q_{k_0}(\Delta) \leq 1$. Namely,
\begin{equation}
|Q_{k_0}(\Delta)| \leq 1, ~~ {\rm for} ~ 0\leq \Delta \leq c/6 .
\label{Chebyshevbound}
\end{equation}
Under the condition $\widehat\Delta> c/6$, the maximum of $Q_{k_0}(\widehat\Delta)$ is achieved by
the degree $(2k_0-1)$ Chebyshev polynomial of the first kind $T_{2k_0-1}(x)$ with $x=\frac{\Delta-c/12}{c/12}$ \cite{ChebyshevPolynomials}. Notably, the polynomial is independent of $\widehat\Delta$.

We were so far optimizing the coefficient of delta function in $\omega_i(\widehat\Delta)$. However, it turns out that the bound we found is also a bound for the integral $\int_{\widehat\Delta}^\infty\omega_i(\Delta)d\Delta$. Indeed, optimizing this integral would replace \eqref{deltacondition} with $P_0(\Delta)\geq 1$ for all $\Delta\geq\widehat\Delta$. It is easy to check that $P_0$ corresponding to the Chebyshev polynomial solution satisfies this stronger constraint as well. This in fact can be generalized to many cases of the form $\int_{\widehat\Delta}^\infty f(\Delta)\omega_i(\Delta)d\Delta$. Therefore,
\begin{equation}
\int_\Delta^\infty  \omega_i (\Delta' ) d\Delta' \leq \frac{1}{1+ T_{2k_0+1}\left(\frac{\Delta-c/12}{c/12}\right)},
\end{equation}
for $\Delta > c/6$.
Similarly, for a general value of $|\tau|<1$, the tail at the threshold $\Delta_\tau$ can be bounded as,
\begin{equation}
 \int_\Delta^\infty \omega_\tau(\Delta') d\Delta' \leq \frac{2}{1+ T_{2k_0-1}\left(\frac{\Delta-\Delta_\tau/2}{\Delta_\tau/2}\right)},
\label{ChebyshevBound2}
\end{equation}
for $\Delta > \Delta_\tau$. To see this, recall the condition \eqref{momentlambda}, which for odd powers of $\lambda$ can rewritten as
\begin{equation}
	\int_{-a}^\infty\lambda^{2k-1}\omega''_\tau(\lambda)d\lambda=0,
\end{equation}
and $a=\mathrm{max}\{\pi|\tau|c/6,\pi c/6|\tau|\}$ and $\omega''_\tau(\lambda)=\frac{1}{2}[w'_\tau(\lambda)+w'_{-1/\tau}(\lambda)]$. Here it is understood that $\omega_\tau(\Delta)=0$ for $\Delta<0$. It is also easy to see the normalization
\begin{equation}
	\int_{-a}^\infty \omega''_\tau(\lambda)d\lambda=1,
\end{equation}
and thus the problem is reduced to $\tau=i$ case. It then follows
\begin{equation}
	\int_{\widehat\lambda}^\infty\omega''_\tau(\lambda)d\lambda\leq \frac{1}{1+T_{2k_0-1}(\widehat\lambda/a)},
\end{equation}
for $\widehat\lambda>a$ which then easily implies the claim.

Note that in the inequality \eqref{ChebyshevBound2} in the denominator is the polynomial which has the largest value for $\Delta>\Delta_\tau$, subject to the requirement of taking values in $[0,2]$ for $0\leq \Delta\leq \Delta_\tau$. In this way, it wins over any polynomial such as \eqref{examplekernel}, especially if one takes $l$ to be asymptotically larger than $\Delta_\tau$ and the degree of $K_c$ smaller than that of the Chebyshev polynomial. More precisely, $K_c$ can be used as $f$ in the aforementioned generalized bound on $\int f(\Delta)\omega_\tau(\Delta)d\Delta$. This justifies truncating the integrals in \eqref{momentlambda}.

\subsection{Crossing Symmetry}

Unlike the case of the partition function in two dimensions, where contributions from Virasoro descendants are subleading in $1/c$,
conformal descendants play an important role in the large $\Delta$ asymptotics in the four-point function (unless one makes a careful choice of the configuration of the four points \cite{Hogervorst:2013sma}). For example, the large $\Delta$
conformal block behaves as $\rho^\Delta$ as we saw in  (\ref{approximateblock}) whereas the contribution
of each local operator is $x^\Delta$, and their difference is not negligible in the large $\Delta$ limit.
 On the other hand, it is easier to derive various bounds on the spectral decomposition
of the four-point function if we use $x^\Delta$. Thus, we will start with the warm-up exercise with the expansion,
\begin{equation}
  G(x) = \int_0^\infty x^{\Delta - 2\Delta_0} g^{(s)}(\Delta) d\Delta,
\label{globalblock}
\end{equation}
where we treat all the local operators (including conformal descendants) independently.

\subsubsection{Reflection and Bounds}

Crossing symmetry $G(x) = G(1-x)$ means $G(x)$ is symmetric under reflection at $x=1/2$, and therefore,,
\begin{equation}
\frac{\partial^{2k+1}}{\partial x^{2k+1}} G(x)_{\big| x=1/2} = 0, ~~ k=0,1,2,\cdots,
\end{equation}
which is equivalent to,
\begin{equation}
  \int_0^\infty \left[ \Delta -2 \Delta_0 \right]^{(2k+1)} \gamma^{(s)}_{1/2}(\Delta) d\Delta = 0.
\label{globalconstraint}
\end{equation}
The bracket symbol $[ \Delta - 2\Delta_0]^{(k)}$ in this subsection is different from the previous one and is the falling Pochhammer symbol,
\begin{align}
  [y]^{(N)} &\equiv x^{N-y} \frac{\partial^N}{\partial x^N}x^y{}_{\big| x=1/2}, \nonumber\\
&= y(y-1)(y-2) \cdots (y-N+1) .
\label{simplebracket}
\end{align}
When $N\ll \sqrt{|y|}$, we can approximate $[y]^{(N)} \sim y^N$. We can then repeat the analysis for the partition function and find that
$\gamma_{1/2}^{(s)}(\Delta) $ is approximately reflection symmetric,
\begin{equation}
\bar\gamma_{1/2}^{(s)}(\Delta) \simeq \bar\gamma_{1/2}^{(s)}(4\Delta_0-\Delta)   .
\label{globalreflectionone}
\end{equation}
 In general,
\begin{equation}
\bar\gamma_{x}^{(s)}(\Delta) \simeq \bar\gamma_{1-x}^{(s)}\left( \frac{2}{x} \Delta_0 - \frac{1-x}{x} \Delta\right)\frac{1-x}{x}   .
\label{globalreflection}
\end{equation}
In particular, $\gamma^{(s)}_x(\Delta < 0)=0$ means,
\begin{equation}
 \bar\gamma^{(s)}_x(\Delta > \Delta_x) \simeq 0,
\end{equation}
where,
\begin{equation}
 \Delta_x = \frac{2}{1-x} \Delta_0.
\label{globalsaddle}
\end{equation}

In this limit, we can also solve the linear optimization problem to find,
\begin{equation}
\int_\Delta^\infty \gamma^{(s)}_{1/2}(\Delta')d\Delta' \leq \frac{1}{1 + T_{2k_0+1}\left(\frac{\Delta - 2\Delta_0}{2\Delta_0}\right)} ,
\end{equation}
for $\Delta>4\Delta_0$.
For general $x$, we can bound $\gamma_x(\Delta)$ for $\Delta > \Delta_x$ by,
\begin{equation}
\int_\Delta^\infty \gamma^{(s)}_x(\Delta') d\Delta' \leq  \frac{2}{1 + T_{2k_0+1}\left(\frac{\Delta - \Delta_x/2}{\Delta_x/2}\right)} .
\end{equation}
The bound is stronger for $x = 1/2$ since $\gamma^{(s)}_{1/2}(\Delta)$ is
invariant under the reflection as in (\ref{globalreflectionone}), while the reflection symmetry for $x \neq 1/2$ relates
 $\gamma^{(s)}_x$ to $\gamma^{(s)}_{1-x}$ as in (\ref{globalreflection}).
The latter bound can be improved in a neighborhood of $x=1/2$.

For $\gamma_x^{(s)}(\Delta)$, we can also derive bounds at finite values of $\Delta$ and $\Delta_0$, without approximating $[y]^{(N)}$ by $y^N$  because of
the simple structure (\ref{simplebracket}) of the bracket symbol.
As we explained in the case of partition function, the problem is to maximize $P(\Delta)$ given by,
\begin{equation}
 P(\Delta) = 1 + \sum_{k=0}^\infty  \left[ \Delta - 2\Delta_0 \right]^{(2k+1)} \lambda_k ,
\end{equation}
at a particular value of $\Delta$ while maintaining $P(\Delta) \geq 0$ for all values of $\Delta$.

However, as we noted before, any $P(\Delta)$ satisfying the constraints will lead to a upper bound on the optimal value of the primal problem. We can use,
\begin{equation}
P(\Delta) = 1 - \frac{[\Delta - 2\Delta_0]^{(2k+1)}}{[-2\Delta_0]^{(2k+1)}},
\end{equation}
as an ansatz for such a $P(\Delta)$. To check that $P(\Delta) \geq 0$, we note that $[-2\Delta_0]^{(2k+1)} < 0$
and $[\Delta - 2\Delta_0]^{(2k+1)} > 0$ for $\Delta - 2\Delta_0 > 2k$, and it is easy to show that
\begin{equation}\left|
\frac{[\Delta - 2\Delta_0]^{(2k+1)}}{[-2\Delta_0]^{(2k+1)}} \right| \leq 1,
\end{equation}
for $\Delta - 2\Delta_0 \leq 2k$, provided $\Delta_0 \geq 1/2$ (this condition can be weakened).
Maximizing this ansatz $P(\Delta)$ at a particular value of $\Delta$ by
using $k$ as a variable gives the bound (after a natural interpolation of the right hand side, which happens not to invalidate the bound),
\begin{equation}
 \int_\Delta^\infty
\gamma^{(s)}_{1/2}(\Delta') d\Delta' \leq
\frac{1}{1
+\frac{\Gamma(\Delta - 2\Delta_0+1)\Gamma(2\Delta_0)}
{\Gamma\left(\frac{\Delta+3}{2}\right)\Gamma\left(\frac{\Delta-1}{2}\right)}}.
\label{finitebound_der}
\end{equation}

The above analysis of the limit $\Delta_0\to\infty$ is easily carried over to the case of conformal blocks. One just has to note that
\begin{align}
	\frac{\partial^n}{\partial x^n} \rho^\Delta x^{-2\Delta_0}&\simeq\left(\frac{\partial\log \rho^\Delta x^{-2\Delta_0}}{\partial x}\right)^n \rho^\Delta x^{-2\Delta_0}\nonumber\\
	&=\left(\frac{\Delta}{x\sqrt{1-x}}-\frac{2\Delta_0}{x}\right)^n\rho^\Delta x^{-2\Delta_0},
\end{align}
to see that a polynomial approximation can be made again. It is then straightforward to derive the corresponding formulas for the conformal block case.

\subsubsection{Cardy formula}
Derivation of the Cardy-like formula \eqref{4ptcardy} for the OPE coefficients is essentially equivalent to the partition function case in \cite{Hartman:2014oaa}. We outline the main steps here.

First, the analogue of light sparse spectrum condition is interpreted using crossing symmetry as, for $x>1/2$,
\begin{equation}
	\log G(x)=-2\Delta_0\log(1-x)+O(1).
\end{equation}
Then, one divides the spectrum into light and heavy parts, $L=[0,2\Delta_0+\epsilon)$ and $H=[2\Delta_0+\epsilon,+\infty)$. Here $\epsilon$ is some fixed positive number, which can be taken exponentially small in $\sqrt{\Delta_0}$. A scaling dimension $\widehat\Delta$ is then picked inside the heavy spectrum and the latter is further split into three parts,
\begin{align}
	H_1=[2\Delta_0+\epsilon,&\widehat\Delta-\delta),\quad H_3=(\widehat\Delta+\delta,+\infty),\\
	&H_2=[\widehat\Delta-\delta,\widehat\Delta+\delta].
\end{align}
Here $\delta$ is some averaging scale which will turn out to be restricted by $\delta\sim \Delta_0^\alpha$, $\alpha\in (1/2,1)$.

The idea is now to show that if $\widehat \Delta=2\Delta_0/(1-x)$, then $G(x)$ is essentially due to contributions from $H_2$, $G\simeq G[H_2]$. To that end, one first bounds $G[H_2]\leq G$, as well as
\begin{equation}
	G[H_2]=\int_{H_2}x^{\Delta-2\Delta_0}g^{(s)}(\Delta)d\Delta\geq x^{\widehat\Delta-2\Delta_0+\delta}\bar{g}^{(s)}_\delta(\widehat\Delta),
\end{equation}
where
\begin{equation}
	\bar{g}^{(s)}_\delta(\widehat\Delta)=\int_{H_2}g^{(s)}(\Delta)d\Delta.
\end{equation}
This leads to an inequality for $\bar{g}^{(s)}(\widehat\Delta)$, which, upon picking an optimal value of $x$, reads for $\widehat\Delta>4\Delta_0$ as
\begin{multline}
	\log\bar g^{(s)}_\delta(\widehat\Delta)\leq -\widehat\Delta \log\left(1-\frac{2\Delta_0}{\widehat\Delta}\right)+
	\\+2\Delta_0\log \left(\frac{\widehat\Delta}{2\Delta_0}-1\right)-\delta\log\left(1-\frac{2\Delta_0}{\widehat\Delta}\right).
	\label{cardyinequality}
\end{multline}
One also gets a different inequality for $2\Delta_0\leq\widehat\Delta\leq 4\Delta_0$. Then one replaces $\delta$ in these inequalities with a new $\delta'$ and takes the latter to be sufficiently small while keeping the $\delta$ in $H_i$ fixed. This allows one to bound the contribution from $H_1$ and $H_3$ up to $\log\Delta_0$ error terms. The contribution from $L$ is also bounded \cite{Hartman:2014oaa}. It then follows that given $\delta\sim \Delta_0^\alpha$, $\alpha\in (1/2,1)$ $H_2$ dominates the 4-point function, and the inequality \eqref{cardyinequality} turns into the equality \eqref{4ptcardy}.


\section{Discussion}

In the present paper we studied implications of modular invariance and crossing symmetry in certain scaling limits.
We have found that all these cases share certain general features, in particular
\begin{enumerate}
\item A truncated set of crossing equations limits to a problem about polynomial moments of the branching ratios. This leads to an approximate duality relation for the branching ratios at crossing symmetric points.
\item The duality relation motivates tail bounds for the integrals of the branching ratios. These bounds are threshold bounds in the sense that they constrain the set of dominant scaling dimensions.
\item ``Sparseness'' of the light spectrum implies universality of the couplings of heavy spectrum. Such theories almost saturate the tail bounds. We discussed this only in two cases, but it is clear that this is a general feature.
\end{enumerate}

These facts have a natural explanation if one assumes that a single saddle point dominates the expansions. Indeed, in this case the location of saddle point can be determined easily by taking appropriate log-derivative of the four-point or partition function. The crossing relation then imposes an equation on this location in a straightforward way. Note, however, that at no point we made such an assumption. In fact, one can assume that several competing saddle points may exist at some points, and in this case our duality relation maps their positions to the crossing symmetric expansion. This happens for example for generalized free field, which at $x=1/2$ exhibits two saddle points -- one at $\Delta=0$ and one at $\Delta=4\Delta_0$ (in scaling blocks). These two saddles are correctly related by the duality relation.

Besides this general features, we have also found features specific for some of the cases, in particular
\begin{enumerate}
\item For scaling block expansion of four point function we were able to use an ansatz incorporating infinitely many derivatives to produce an exponentially decaying tail bound. This bound is a strict inequality valid without taking any limit whatsoever.
\item For the large spacetime dimension limit of the conformal block expansion, we were able to see a manifestation of unitarity bound for external scalars without the use of the free scalar equation of motion.
\end{enumerate}

Most of our results used some kind of a limit, and thus are not applicable to the bootstrap of light operators. However, one may hope that some qualitative features also carry over to the case of light operators, and thus may provide useful intuition. Let us discuss possible implications for numerical analysis.

In some cases, the four-point amplitude $G(x)$ is dominated by operators near the saddle point $\Delta(x)$. This observation
may have applications to numerical bootstrap methods. which often employ derivatives of the crossing relation at $x=1/2$.
This mostly probes operators near the saddle point $\Delta(1/2)$.
To learn about the other parts of the spectrum, apart from taking more and more derivatives at $x=1/2$, one may consider the crossing relation at different values of $x$.
In the case of scaling blocks, it is natural to expect that $O(1)$ changes in $1/(1-x)$ result in $O(\Delta_0)$ changes in $\Delta(x)$ and that the width of saddle point is on the order of $\sqrt\Delta_0$.
Therefore, in order to have the spectrum up to $\Delta=\Lambda$ evenly covered, one may use the bootstrap equation at $O(\Lambda\sqrt{\Delta_0})$ points $x$ so that $1/(1-x)$ is distributed evenly
with spacing of the order of $O(1/\sqrt{\Delta_0})$.

Another observation is that gaps in OPE spectrum can render the parts of spectrum symmetric to them difficult to study. An example is the generalized free field four point function, which is
\begin{equation}
	G(x)=\frac{1}{x^{2\Delta_0}}+\frac{1}{(1-x)^{2\Delta_0}}+1.
\end{equation}
It can be easily seen to be dominated by the vacuum term $x^{-2\Delta_0}$ for $x<1/2$. As discussed above, this forces a discontinuity in $\Delta(x)$ at $x=1/2$, with $\Delta(1/2+0)=4\Delta_0$. In this theory there are no operators in the interval $(0,2\Delta_0)$, but there are operators in $[2\Delta_0,4\Delta_0)$, which by the approximate symmetry never dominate the four-point function.

\vspace{0.5cm}
\noindent
\textbf{Note added:}
Toward completion of this manuscript, the paper \cite{Chang:2015qfa} appeared, which uses an idea similar to that developed in this paper to study the semi-classical limit of conformal field theory in two dimensions.

We have also been informed on the forthcoming
paper \cite{RychkovYvernay}, which
improves the results of \cite{Pappadopulo:2012jk} for the convergence of conformal block expansion. The new paper introduces a surprisingly simple expression for conformal blocks on the real line in 3D, and then uses it to transform the bounds of \cite{Pappadopulo:2012jk} for ``descendant'' expansion \eqref{fourpointlimit} to a better asymptotic bound for the exact conformal block expansion. It would be interesting to try to combine their approach with the one presented in this paper to obtain stronger non-asymptotic bounds.

Preliminary versions of this work were presented at the Eurostrings conference at Cambridge University in March 2015, at the conference ``(Mock) Modularity, Moonshine and Strings Theory'' at Perimeter Institute in April 2016, and at the Aspen Center for Physics in July 2016.

\section*{Acknowledgments}

We thank S.~El-Showk, N.~Hunter-Jones, C.~Keller, Z.~Komargodski, Y.~Nakayama, S.~Rychkov, D.~Simmons-Duffin, B.~Stoica, and W.~Yan for discussion.
We also thank S.~Rychkov and D.~Simmons-Duffin for their comments on the draft
of this paper.
This work is supported in part by U.S.\ DOE grant DE-SC0011632,
by the WPI Initiative of MEXT, by JSPS KAKENHI Grant Numbers C-26400240 and 15H05895,
by the Simons Investigator Award, and by
the Walter Burke Institute for Theoretical Physics
and the Moore Center for Theoretical Cosmology and
Physics at Caltech.
We thank the hospitality of the Institute for Advanced Study, where HO is Director's Visiting Professor.
HO also thanks the Aspen Center for Physics and
the Simons Center for Geometry and Physics, where parts of this work were done.

\appendix

\section{Asymptotic form of conformal blocks on the diagonal $x=\bar x$}
\label{app:diagonalLimit}

In \cite{Hogervorst:2013kva} a fourth-order differential equation was derived
for the conformal blocks in $\dspace$ dimensions on the diagonal $x=\bar x$. The
derivation is based on combining the quadratic and quartic Casimir equations.

The equation has the following form
\begin{equation}
	D_4 f(x)=0,\label{eq:diagonaleq}
\end{equation}
where the differential operator $D_4$ is defined below and $f(x)=F_{\Delta,\ell}(x=\bar x)$. This equation is equipped with the boundary condition
\begin{equation}
	f(x)\sim x^\Delta,\,x\to 0,
\end{equation}
which also fixes our normalization.

Another way to phrase our normalization is to say that the conformal block for complex $x$ is given by
\begin{equation}
	F_{\Delta,\ell}(x,\bar x)\sim r^\Delta \frac{C_\ell^{(\epsilon)}(\cos\theta)}{C_\ell^{(\epsilon)}(1)},
\end{equation}
where $r=|x|\to 0$, $\theta=\arg x$, $\epsilon=d/2-1$ and $C_\ell^{(\epsilon)}$ is the Gegenbauer polynomial.

The operator $D_4$ is given by
\begin{align}
	D_4=&(x-1)^3x^4\frac{d^4}{dx^4}+\sum_{r=2}^3(x-1)^{r-1}p_r(x)x^r\frac{d^r}{dx^r}+\nonumber\\
	&+\sum_{r=0}^1p_r(x)x^r\frac{d^r}{dx^r},\label{eq:diagonalD4}
\end{align}
where $p_3,\,p_2,\,p_1,\,p_0$ are known \cite{Hogervorst:2013kva} polynomials in $x$ of degrees $1,\,2,\,3,\,3$ respectively, whose coefficients depend on the differences between external operator scaling dimensions, which we set to 0 (then $p_0$ is degree $2$), as well as on the spin and scaling dimension of the intermediate operator. The dependence of $p_r$ on $\Delta$ and $\ell$ is through the quadratic and quartic Casimir invariants $c_2$ and $c_4$,
\begin{align}
	c_2=&\tfrac{1}{2}\left[\ell(\ell+2\epsilon)+\Delta(\Delta-2-2\epsilon)\right],\\
	c_4=&\ell(\ell+2\epsilon)(\Delta-1)(\Delta-1-2\epsilon).
\end{align}

We will be considering the double-scaling limit with $\lambda=\ell/\Delta$ fixed and $\Delta$ large. We will do so in order to allow for large angular momenta. Our results will turn out to be applicable to small angular momenta as well by setting the ratio $\lambda$ to be $0$.

\begin{figure*}[t]
	\centering
	\begin{subfigure}[b]{0.45\textwidth}
	\includegraphics[width=\textwidth]{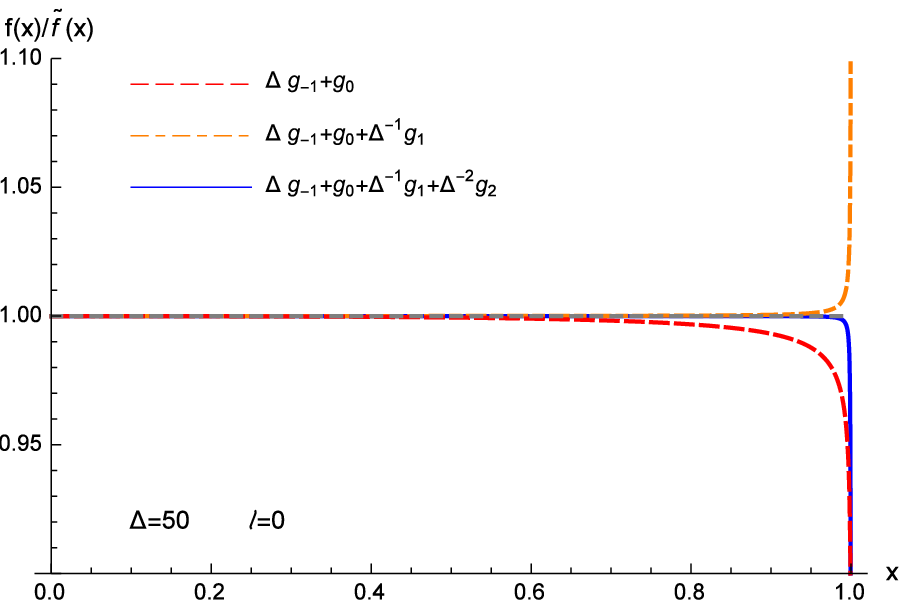}
	\caption{$\Delta=50,\,\,\ell=0$}
	\label{fig:block50_0}
	\end{subfigure}
	~
	\begin{subfigure}[b]{0.45\textwidth}
	\includegraphics[width=\textwidth]{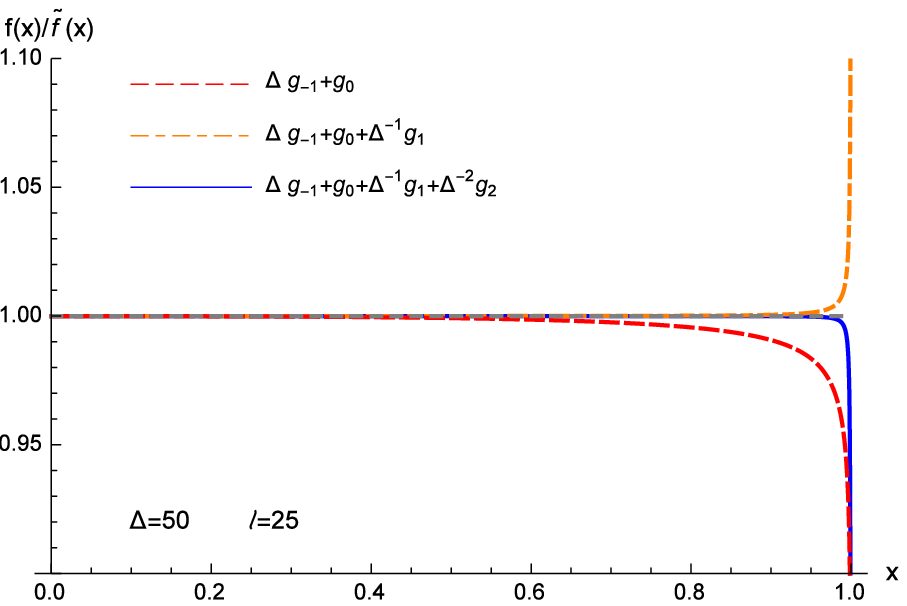}
	\caption{$\Delta=50,\,\,\ell=25$}
	\label{fig:block50_25}
	\end{subfigure}
	\caption{Plots of $f(x)/\tilde{f}(x)$ for different orders of approximation and values of angular momenta in four dimensions.}
	\label{fig:block_comp}
\end{figure*}

With this scaling assumed, $c_2\propto\Delta^2$ and $c_4\propto\Delta^4$. The polynomials have the leading behavior
\begin{align}
	p_0\simeq& c_4(x-1),\\
	p_1\simeq& c_2(1-2\epsilon)x^2+c_2(1+6\epsilon)x-2c_2(1+2\epsilon),\\
	p_2\simeq& 2c_2(x-1),\\
	p_3=& O(1).	
\end{align}
We would like to see whether there is a WKB-like solution of the form $f(x)\sim e^{\Delta g(x)}$, where $g(x)=O(1)$. It is easy to see that the leading power of $\Delta$ produced by action of \eqref{eq:diagonalD4} on such a solution will be $\Delta^4$ since each derivative produces a power of $\Delta$, and the polynomials $p_r$ have scaling $\Delta^k$ with $k\leq 4-r$. We see that in the leading $\Delta^4$ order only $p_0$ and $p_2$ appear. This results in the equation for $g$ (for $x<1$)
\begin{equation}
	\left[\sqrt{1-x}xg'(x)\right]^4-2\frac{c_2}{\Delta^2}\left[\sqrt{1-x}xg'(x)\right]^2+\frac{c_4}{\Delta^4}=0
\end{equation}
Here we are only allowed to keep the leading terms in the Casimir invariants. We then find the following solutions,
\begin{equation}
	\left[\sqrt{1-x}xg'(x)\right]^2=1\,\quad \left[\sqrt{1-x}xg'(x)\right]^2=\frac{\ell^2}{\Delta^2}.
\end{equation}
With our boundary condition we are interested in $\sqrt{1-x}xg'(x)=1$ which produces
\begin{equation}
	g(x)=\log\rho,
\end{equation}
where
\begin{equation}
	\rho=\frac{4x}{(1+\sqrt{1-x})^2}.
\end{equation}
We thus find that $\log f(x)=\log \rho^\Delta+O(1)$ is a solution. We can perform the analysis more systematically by substituting $f(x)=e^{G(x)}$ in \eqref{eq:diagonaleq} and looking for $g$ in the form
\begin{equation}
	G(x)=\Delta g_{-1}(x)+g_0(x)+\frac{1}{\Delta}g_1(x)+\frac{1}{\Delta^2}g_2(x)+\ldots
\end{equation}
Then we will be able to solve the resulting equation order by order in $\Delta$. We already found $g_{-1}(x)=\log \rho$. The next order gives
\begin{equation}
	f(x)=\left(1-\frac{\rho^2}{16}\right)^{-\epsilon-1}\rho^\Delta e^{O(\frac{1}{\Delta})},\label{eq:genericasympt}
\end{equation}
this not depending on whether we scale $\ell$ with $\Delta$ or not.

\begin{figure}[t]
	\centering
	\includegraphics[width=\columnwidth]{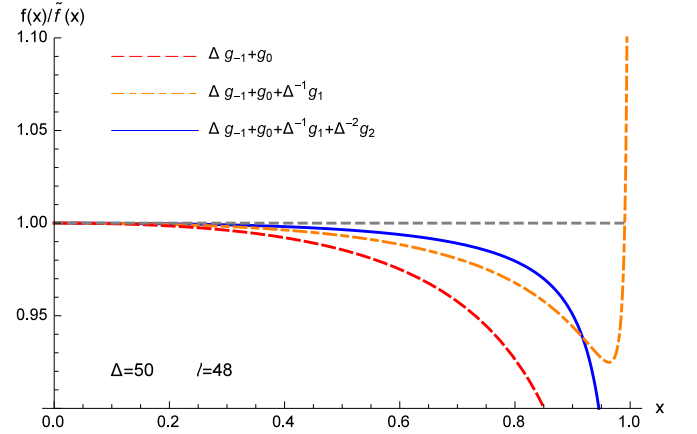}
	\caption{Plot of $f(x)/\tilde{f}(x)$ for different orders of approximation and the value of angular momenta at the unitarity bound in four dimensions.}
	\label{fig:block50_48}
\end{figure}

Order by order we have
\begin{align}
	g_{-1}&=\log\rho,\\
	g_0&=-(1+\epsilon)\log\left(1-\frac{\rho^2}{16}\right),\\
	g_1&=\frac{\rho^2}{16}\frac{1}{1-\frac{\rho^2}{16}}\frac{(1+\epsilon-\epsilon^2)\Delta^2+\epsilon(\epsilon-1)\ell^2}{\Delta^2-\ell^2},\\
	\ldots
\end{align}
The higher order terms get more messy, but are not hard to compute in principle. We can see that $g_2$ contains a negative power of $\Delta^2-\ell^2$, which scales as $\Delta^2$ and is supposed to be canceling $\Delta^2$ scaling in numerator. This means that applicability of our expansion is limited to the region where $\Delta^2-\ell^2$ is not too small. Higher order terms have higher powers of $\Delta^2-\ell^2$ in denominators. We also observe that that the subleading terms become singular in the limit $\rho\to 4$ corresponding to $x\to 1$. Therefore, the above approximation works as an asymptotic expansion when
\begin{enumerate}
\item $\frac{\vert\Delta-\ell\vert}{\Delta}$ is greater than some fixed positive number
\item $x\leq x_0$, where $x_0<1$ and is fixed.
\end{enumerate}
We compare the proposed expansion with the exact conformal block in four dimensions in Fig.~\ref{fig:block_comp}. There $\tilde{f}$ is the approximate conformal block given by our expansion. We include various numbers of terms in the expansion, up to $\Delta^{-2}g_2$. We see that the approximation works almost equally well for scalar (Fig.~\ref{fig:block50_0}) and large-spin (Fig.~\ref{fig:block50_25}) operators. We also observe the promised singularity at $x=1$. See Fig.~\ref{fig:block50_48} for comparison at the unitarity bound.

We can get an understanding of how the conformal block behaves when $\ell\to\Delta$ independently of the above thanks to the decoupling of large numbers of descendants for leading twist operators \cite{Hogervorst:2013sma}. The unitarity limits the maximal spin of an operator to be $\ell=\Delta-d+2=\Delta-2\epsilon$. It is shown in \cite{Hogervorst:2013sma} that for the maximal allowed spin the conformal block on the diagonal $x=\bar x$ can be expressed as
\begin{align}
	f(x)&=\sum_{n=0}^{\infty}\frac{(\ell+\epsilon)_n(\ell+2\epsilon)_n}{n!(2\ell+2\epsilon)_n}x^{\Delta+n}=\nonumber\\
	&=x^\Delta{}_2F_1(\Delta-\epsilon,\Delta;2\Delta-2\epsilon;x).
\end{align}

We can then use the standard representation
\begin{equation}
	B(b,c-d){}_2F_1(a,b;c;x)=\int_0^1 x^{b-1}(1-x)^{c-b-1}(1-xx)^{-a}dx
\end{equation}
to compute the asymptotic expansion of the hypergeometric function by saddle-point method. This leads to
\begin{equation}
	f(x)=\left(1-\frac{\rho^2}{16}\right)^{-\epsilon-1/2}\rho^\Delta\left[1+O\left(\frac{1}{\Delta}\right)\right],
	\label{eq:leadtwistasympt}
\end{equation}
valid at the unitarity bound $\ell=\Delta-2\epsilon$.

We see that for most values of $\ell$, the conformal block can be well approximated by \eqref{eq:genericasympt}. This approximation breaks down as we approach the unitarity bound $\ell=\Delta-2\epsilon$ due to higher-order terms becoming large. However, at the exact unitarity bound the formula \eqref{eq:leadtwistasympt} is valid. The two formulas are compared in Fig.~\ref{fig:block50_48}.

The most important part of the conformal block for us is the factor $\rho^\Delta$, which encapsulates the leading behavior in the limit of large $\Delta$. We see that in all regimes this factor is present and is not modified.

\section{Asymptotic form of conformal blocks on the diagonal \(x=\bar x\) in the large-dimension limit}
\label{app:doublescale}
The previous derivation holds for fixed number of spacetime dimensions and large conformal dimensions of the intermediate state; it therefore captures the behavior of states very far from the unitarity bound. If we additionally adjust the number of spacetime dimensions \(d\), however, we can take analytic approximations that capture the behavior of states close to the unitarity bound. Such a limit was already described in \cite{Fitzpatrick:2013sya}, where the authors derive an expression for the conformal block in the scaling limit
\begin{gather}
d\to\infty,\qquad\Delta\to\infty,\\
\alpha=2-d/\Delta\,\,\text{  fixed}.
\end{gather}
If one takes \(\dspace\to\infty\), then the unitarity bound means that \(\Delta\) and \(\ell\) must scale as well.

To express the conformal block in this limit, define
\begin{align}\label{eq:yDefinition}
y_+&=\frac{x\bar x}{(1+|1-x|)^2}=\frac{x^2}{(2-x)^2},\\
y_-&=\frac{x\bar x}{(1-|1-x|)^2}=1,
\end{align}
where the second equality in each line holds on the real line $x=\bar x$.
The conformal block then becomes, in normalization of \cite{Fitzpatrick:2013sya}
\begin{align}
F_{\Delta,\ell}(x)&\approx\frac{2^{\Delta+\ell}}{\sqrt{y_--y_+}}
A_\Delta(y_+)A_{1-\ell}(y_-)=\\
=&\NN_\ell 2^\Delta\sqrt{\frac{y_-}{y_--y_+}}A_\Delta(y_+)C_\ell^{\left(\frac{d-2}{2}\right)}\left(y_-^{-\half}\right),
\end{align}
where
\begin{align}
	A_\beta(x)&=x^{\beta/2}\hypgeom\del{\frac{\beta-1}{2},\frac{\beta}{2},\beta-\frac{\dspace-2}{2};x},\\
	\NN_\ell&=\frac{\Gamma(\ell+1)\Gamma\left(\frac{d-2}{2}\right)}{\Gamma\left(\ell+\frac{d-2}{2}\right)},
\end{align}
and $C^{(\lambda)}_n(x)$ are the Gegenbauer polynomials.

Notice that the spin dependence factorizes. In particular, when $y_-=1$, spin-dependent factors carry no dependence on $y_+$. This immediately implies that in the normalization of this paper the block has no $\ell$-dependence on real line $x=\bar x$. In fact, we have in our normalization
\begin{equation}
	F_{\Delta,\ell}(x)\approx\frac{(4y_+)^{\Delta/2}}{\sqrt{1-y_+}}\hypgeom\del{\frac{\Delta-1}{2},\frac{\Delta}{2},\Delta-\frac{\dspace-2}{2};y_+}.
\end{equation}

The saddle-point approximation for the hypergeometric function gives
\begin{equation}
\lim_{c\to\infty}\frac{\log\hypgeom(c,c;\alpha c;y)}c=\log\frac{\alpha t_0}{1-yt_0}\left(\frac{\alpha(1-t_0)}{\alpha-1}\right)^{\alpha-1},
\end{equation}
where
\begin{equation}
t_0(\alpha,y)=\frac{\alpha-\sqrt{\alpha^2+4(1-\alpha)y}}{2(\alpha-1)y}.
\end{equation}
Therefore, up to $O(1)$ factors we have
\begin{equation}
F_{\Delta,\ell}(x)\sim\del{\frac{4\alpha y_+t_0}{1-y_+t_0}\left(\frac{\alpha(1-t_0)}{\alpha-1}\right)^{\alpha-1}}^{\Delta/2}.
\end{equation}
A simple computation then gives
\begin{equation}
	\frac{\partial\log F_{\Delta,\ell}}{\partial x}=
	\frac{\dspace(2-x)}{4x(1-x)}\left(1+\sqrt{1+\frac{16(1-x)}{(2-x)^2}\delta(\delta-1)}\right),\label{eq:doublescalelog}
\end{equation}
where $\delta=\Delta/\dspace$.
This is obviously a non-decreasing function of $\Delta$, which for $\delta\gg 1$ asymptotes to
\begin{equation}
	\dspace^{-1}\frac{\partial\log F_{\Delta,\ell}}{\partial x}\approx
	\frac{\delta-1/2}{x\sqrt{1-x}}+\frac{2-x}{4x(1-x)}.
\end{equation}

Note that for $\dez\gg\dspace$ we expect $\Delta\gg\dspace$ to be important and thus $\delta\gg1$, so we regain from this expression the previously discussed case of large $\dez$.

Note that the image of $\beta\in[\half,\infty)$ under $\dspace^{-1}\partial\log F_{\Delta,\ell}/\partial x$ is
\begin{equation}
	[\half,\infty)\mapsto [(2x(1-x))^{-1},\infty),
\end{equation}
as used in the main text.



\begin{thebibliography}{99}

\bibitem{Dolan:2000ut}
  F.~A.~Dolan and H.~Osborn,
  Nucl.\ Phys.\ B {\bf 599}, 459 (2001)
  [hep-th/0011040].

\bibitem{Cardy:1986ie}
  J.~L.~Cardy,
  Nucl.\ Phys.\ B {\bf 270}, 186 (1986).

\bibitem{Ferrara:1973yt}
  S.~Ferrara, A.~F.~Grillo and R.~Gatto,
  Annals Phys.\  {\bf 76}, 161 (1973).


\bibitem{Polyakov:1974gs}
  A.~M.~Polyakov,
  Zh.\ Eksp.\ Teor.\ Fiz.\  {\bf 66}, 23 (1974).

\bibitem{Belavin:1984vu}
  A.~A.~Belavin, A.~M.~Polyakov and A.~B.~Zamolodchikov,
  Nucl.\ Phys.\ B {\bf 241}, 333 (1984).

\bibitem{Rattazzi:2008pe}
  R.~Rattazzi, V.~S.~Rychkov, E.~Tonni and A.~Vichi,
  JHEP {\bf 0812}, 031 (2008)
  [arXiv:0807.0004 [hep-th]].

\bibitem{Pappadopulo:2012jk}
  D.~Pappadopulo, S.~Rychkov, J.~Espin and R.~Rattazzi,
  Phys.\ Rev.\ D {\bf 86}, 105043 (2012)
  [arXiv:1208.6449 [hep-th]].


\bibitem{Hartman:2014oaa}
  T.~Hartman, C.~A.~Keller and B.~Stoica,
  JHEP {\bf 1409}, 118 (2014)
  [arXiv:1405.5137 [hep-th]].


\bibitem{Hogervorst:2013kva}
  M.~Hogervorst, H.~Osborn and S.~Rychkov,
  JHEP {\bf 1308}, 014 (2013)
  [arXiv:1305.1321 [hep-th]].

\bibitem{Hogervorst:2013sma}
  M.~Hogervorst and S.~Rychkov,
  Phys.\ Rev.\ D {\bf 87}, 106004 (2013)
  [arXiv:1303.1111 [hep-th]].
\bibitem{Saff}
  E.~B.~Saff and V.~Totik,
  Lond.\ Math.\ Soc. (2) {\bf 39}, 487 (1989).


\bibitem{ConvexOptimization}
  S.~Boyd and L.~Vandenberghe,
{\it Convex Optimization}
(Cambridge University Press, 2004).

\bibitem{Krein}
  M.~G.~Krein and A.~A.~Nudel'man,
{\it The Markov Moment Problem and Extremal Problems,}
Translations of Mathematical Monographs, Vol. 50
(American Mathematical Society, 1977).

\bibitem{ChebyshevPolynomials}
 T.~J.~Rivlin, {\it The Chebyshev Polynomials} (A Wiley-Interscience publication, 1974).

\bibitem{Chang:2015qfa}
  C.~M.~Chang and Y.~H.~Lin,
  [arXiv:1510.02464 [hep-th]].

\bibitem{RychkovYvernay}
  S.~Rychkov and P.~Yvernay, to appear.

\bibitem{Fitzpatrick:2013sya}
  A.~L.~Fitzpatrick, J.~Kaplan and D.~Poland,
  JHEP {\bf 1308}, 107 (2013)
  [arXiv:1305.0004 [hep-th]].


\end{thebibliography}
\end{document}